\documentclass[%
 reprint,
 amsmath,amssymb,
 aps,prx,
 twocolumn,
 showpacs,
 superscriptaddress,
 floatfix,
 showkeys,
 nofootinbib,
 reprint,
 longbibliography,]
 {revtex4-2}

\usepackage{graphicx}
\usepackage{dcolumn}
\usepackage{bm}
\usepackage{amsmath,amssymb}
\usepackage{hyperref}
\usepackage{natbib}
\usepackage{xcolor}
\usepackage[normalem]{ulem}
\usepackage{booktabs}
\usepackage{tabularx}
\usepackage[english]{babel}
\usepackage[utf8]{inputenc}
\usepackage{enumerate}
\usepackage{newunicodechar}
\usepackage{lipsum}
\usepackage{cleveref}
\usepackage{ulem}
\usepackage{ragged2e} 
\usepackage{xr}       
\usepackage[normalem]{ulem} 

\usepackage[font=small,format=plain,labelfont=bf,
singlelinecheck=false,justification=raggedright,
labelsep=space,figurename=Fig.]{caption}
\makeatletter
\renewcommand{\NAT@open}{}  
\renewcommand{\NAT@close}{} 
\renewcommand{\NAT@cite}[1]{\textsuperscript{\normalfont{#1}}} 
\makeatother

\begin{document}
\title{Skyrmions in 2D chiral magnets with noncollinear ground states stabilized by higher-order interactions}

\author{Mathews Benny}
\thanks{These authors contributed equally to this work}
\affiliation{School of Physics, Indian Institute of Science Education and Research Thiruvananthapuram, Thiruvananthapuram, Kerala 695551, India}

\author{Moinak Ghosh}
\thanks{These authors contributed equally to this work}
\affiliation{Center for High-Performance Computing, Indian Institute of Science Education and Research Thiruvananthapuram, Thiruvananthapuram, Kerala 695551, India}

\author{Moritz A. Goerzen}
\affiliation{CEMES, Universit\'e de Toulouse, CNRS, 29 rue Jeanne Marvig, F-31055 Toulouse, France}
\affiliation{Institute of Theoretical Physics and Astrophysics, Christian-Albrechts-Universit$\ddot{a}$t zu Kiel, Leibnizstrasse 15, 24098 Kiel, Germany}

\author{Bjarne Beyer}
\affiliation{Institute of Theoretical Physics and Astrophysics, Christian-Albrechts-Universit$\ddot{a}$t zu Kiel, Leibnizstrasse 15, 24098 Kiel, Germany}

\author{Hendrik Schrautzer}
\affiliation{Science Institute and Faculty of Physical Sciences, University of Iceland, VR-III, 107 Reykjav\'{i}k, Iceland}

\author{Stefan Heinze}
\email{heinze@physik.uni-kiel.de}
\affiliation{Institute of Theoretical Physics and Astrophysics, Christian-Albrechts-Universit$\ddot{a}$t zu Kiel, Leibnizstrasse 15, 24098 Kiel, Germany}
\affiliation{Kiel Nano, Surface, and Interface Science (KiNSIS), Christian-Albrechts-Universit$\ddot{a}$t zu Kiel, Christian-Albrechts-Platz 4, 24118 Kiel, Germany}

\author{Souvik Paul}
\email{souvikpaul@iisertvm.ac.in}
\affiliation{School of Physics, Indian Institute of Science Education and Research Thiruvananthapuram, Thiruvananthapuram, Kerala 695551, India}
\affiliation{Center for High-Performance Computing, Indian Institute of Science Education and Research Thiruvananthapuram, Thiruvananthapuram, Kerala 695551, India}

\date{\today}

\begin{abstract}
Magnetic skyrmions are intriguing topological spin textures that have attracted great attention due to their potential for future spintronic devices. Skyrmions have so far been explored in different magnetic materials, such as ferromagnets, antiferromagnets, and ferrimagnets. Here, we propose a new type of unconventional skyrmions stabilized in noncollinear magnets. Using first-principles calculations and atomistic spin simulations, we demonstrate that a noncollinear ground state can be stabilized in Rh/Co and Pd/Co atomic bilayers on the Re(0001) surface by four spin exchange interactions, although Co -- a material often used in applications -- is a prototypical ferromagnet with strong pairwise exchange interaction. We further show that unconventional skyrmion lattices and isolated skyrmions can emerge on this noncollinear magnetic background. Transition-state theory calculations reveal that these metastable skyrmions are protected by large energy barriers, suggesting that they could be observed in experiments. These unconventional types of skyrmions in noncollinear magnets might open new possibilities for topological spin transport or magnet-superconductor hybrid systems.
\end{abstract}

\maketitle
\section*{}
\noindent{\textbf{\MakeUppercase{Introduction}}\\
Due to the intriguing topological and dynamical properties, magnetic skyrmions~\textsuperscript{\cite{nagaosa13,Bogdanov1989}} 
-- localized, stable spin structures with diameters down to the nanometer scale -- are of fundamental interest and hold great promise for spintronic applications~\textsuperscript{\cite{tomasello14,zhang2015}},
as well as for quantum~\textsuperscript{\cite{psaroudaki2021,psaroudaki2022}} and neuromorphic computing~\textsuperscript{\cite{grollier2020,song2020}}. These whirling topological spin structures were first reported in bulk materials~\textsuperscript{\cite{Muehlbauer2009,neubauer2009}}
with broken inversion symmetry as metastable quasiparticles on a ferromagnetic (FM) background. Skyrmions were also observed at transition-metal (TM) interfaces~\textsuperscript{\cite{Heinze2011,Romming2013,Romming2015,meyer19}}
and in multilayer heterostructures~\textsuperscript{\cite{Boulle2016,Moreau-Luchaire2016,Woo2016}}
These material systems are ideally suited for practical applications, since they can be directly integrated with existing spintronic technologies, and their properties can be altered easily by varying the chemical composition and interfacial structure~\textsuperscript{\cite{Dupe2016}}. 

Recently, skyrmions have also been realized in two-dimensional (2D) van der Waals magnets~\textsuperscript{\cite{han2019,ding2020,wu2020}},
synthetic antiferromagnets~\textsuperscript{\cite{legrand2020,he2024}}, and 
ferrimagnets~\textsuperscript{\cite{Woo2018a,Woo2018b,Hirata2019,Xu2023}} and have been proposed to appear in intrinsic antiferromagnetic materials~\textsuperscript{\cite{zhang2016,barker2016,amal2022}}. However, there are also 2D material systems in which a noncollinear ground state is stabilized by higher-order exchange interactions~\textsuperscript{\cite{Yoshida2012,romming18,spethmann2020}}
This raises the intriguing question of whether similar types of metastable isolated skyrmions can even emerge on such complex magnetic ground states. Skyrmions on a noncollinear magnetic background have recently been reported experimentally in a bulk material~\textsuperscript{\cite{chakrabartty2025}}. Based on first-principles calculations, meronic spin structures have also been predicted to be stable on an in-plane N\'eel state in ultrathin Mn films on the Ir(111) surface~\textsuperscript{\cite{amal2023}}. However, realizations of skyrmions on a noncollinear magnetic background of the 2D materials or ultrathin film systems have up to now been elusive.

The stability of skyrmions in typical 2D chiral magnets results from competition among the Heisenberg pairwise exchange interaction, the Dzyaloshinskii-Moriya interaction (DMI)~\textsuperscript{\cite{Dzyaloshinskii1957,Moriya1960}},
and the magnetocrystalline anisotropy energy (MAE), which comprise the standard spin model. The DMI arises from spin-orbit coupling in systems with broken structural inversion symmetry, i.e., at an interface. It is responsible for the chiral nature of skyrmions~\textsuperscript{\cite{bode2007}} and contributes to their stability. Exchange frustration can lead to a large enhancement of skyrmion stability~\textsuperscript{\cite{malottki2017a}}. In TM films, multi-spin interactions beyond the pairwise Heisenberg exchange, such as the biquadratic, the three-site four spin, and the four-site four spin interactions, can arise from fourth-order perturbation theory of the Hubbard model~\textsuperscript{\cite{takahashi77,macD88,hoffmann20}}. These higher-order exchange interactions (HOI) can contribute significantly to the energy barriers of skyrmions in ultrathin TM films and stabilize them at interfaces or in 2D films even in the absence of DMI~\textsuperscript{\cite{paulhoi}}. Various complex magnetic ground states have been reported to be stabilized by HOI in Fe~\textsuperscript{\cite{Heinze2011,romming18,paul2020b,li2020,gutzeit2021}} and Mn~\textsuperscript{\cite{Kurz2001,Yoshida2012,spethmann2020}} based ultrathin films. However, studies on the stabilization of topological skyrmions due to HOI are still rare~\textsuperscript{\cite{schrautzer2025impact}}.

Here, we propose a new class of skyrmions that can emerge spontaneously on the complex spin structure of a noncollinear magnet. We predict that these unconventional types of skyrmions can occur in ultrathin film systems, such as Pd/Co and Rh/Co atomic bilayers on the Re(0001) surface, and can be experimentally revealed by spin-polarized scanning tunneling microscopy (SP-STM). Our prediction is based on atomistic spin simulations for an extended spin model in which all interactions are obtained from first-principles calculations. The noncollinear ground state -- which arises at zero and finite magnetic fields for Pd/Co and Rh/Co bilayers, respectively -- is stabilized by four-spin interactions, which are neglected in the standard spin model. Surprisingly, these unconventional skyrmions emerge at interfaces containing Co, which is a prototypical ferromagnet dominated by strong pairwise exchange interactions.

We demonstrate that the crucial difference compared to known material systems is the sign and magnitude of the four-site and three-site four spin interactions. These multi-spin interactions destabilize the FM state and promote a novel type of noncollinear ground state. Isolated skyrmions are stabilized on this noncollinear spin background due to DMI and frustrated exchange interaction, as demonstrated via the minimum energy path calculations using the geodesic nudged elastic band (GNEB) method. The large energy barriers found for these isolated skyrmions with diameters on the order of 10 nm suggest that their stability is comparable to skyrmions on collinear FM backgrounds observed previously at transition-metal interfaces. The skyrmions predicted here extend the existing family of topological quasiparticles beyond the conventional FM, antiferromagnetic, and ferrimagnetic backgrounds with the potential of novel dynamical and topological transport properties.\\

\noindent{\textbf{\MakeUppercase{Results}}\\
$\textbf{Magnetic phase diagram.}$ We have studied the relaxation and energetics of various spin structures in ultrathin magnetic films using the following atomistic spin Hamiltonian~\textsuperscript{\cite{malottki2017a,som18,dupe2014,paul2020b,paulhoi,paul2022,paul2024}}
\begin{widetext}
\begin{gather} \label{eq:hamiltonian}
\mathcal{H}=- \sum_{ij} J_{ij} (\textbf{m}_{i} \cdot \textbf{m}_{j}) - \sum_{ij} \boldsymbol{D}_{ij} \cdot (\textbf{m}_{i}\times\textbf{m}_{j}) - \sum_{i} K(m^{z}_{i})^2 - \sum_{i} \mu_{s} \textbf{B} \cdot \textbf{m}_{i} - B_{1} \sum_{<ij>} (\textbf{m}_{i} \cdot \textbf{m}_{j})^2 \nonumber \\ -2~Y_{1} \sum_{<ijk>} (\textbf{m}_{i} \cdot \textbf{m}_{j}) (\textbf{m}_{j} \cdot \textbf{m}_{k}) - K_{1} \sum_{<ijkl>} (\textbf{m}_{i} \cdot \textbf{m}_{j}) (\textbf{m}_{k} \cdot \textbf{m}_{l}) +(\textbf{m}_{i} \cdot \textbf{m}_{l}) (\textbf{m}_{j} \cdot \textbf{m}_{k}) -(\textbf{m}_{i} \cdot \textbf{m}_{k}) (\textbf{m}_{j} \cdot \textbf{m}_{l})  
\end{gather}
\end{widetext}

\begin{figure*}[htbp!]
	\centering
	\includegraphics[scale=1.0]{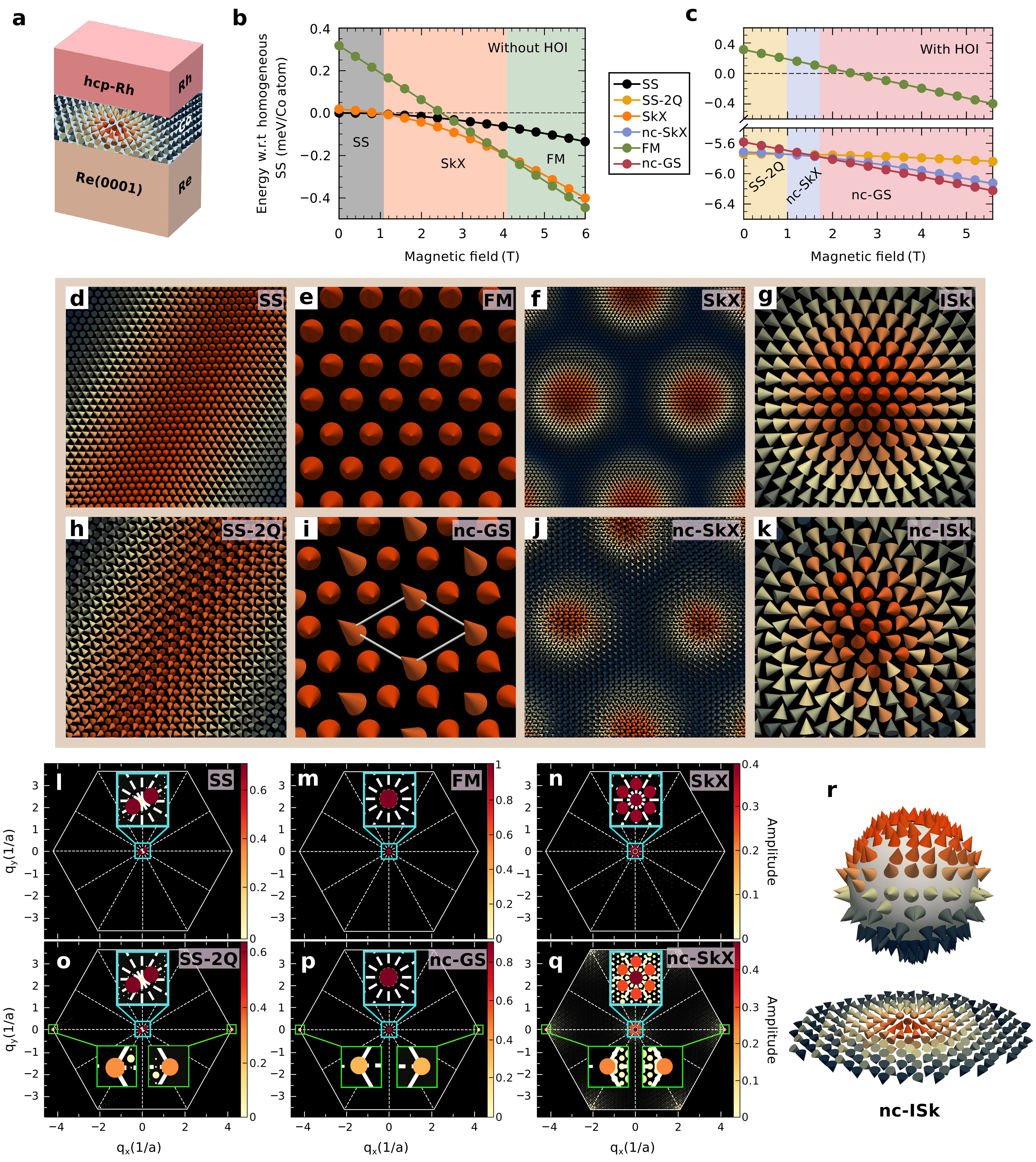}
	\caption{\justifying \textbf{Magnetic phase diagram, spin structures, and Fourier transforms of Rh/Co/Re(0001) including and excluding HOI.} \textbf{a} Illustration of an isolated skyrmion in the Co layer of Rh/Co/Re(0001). \textbf{b,c} Zero temperature phase diagram of Rh/Co/Re(0001) obtained by neglecting HOI and including all interactions, i.e., including HOI, respectively. Energies of spin spirals (SS, black circles and line), skyrmion lattice on the FM background (SkX, orange circles and lines), FM state (green circles and lines), excluding HOI, and energies of the spin spirals (SS-2$Q$, yellow circles and line), noncollinear ground state (nc-GS, red circles and lines), skyrmion lattice on the same noncollinear background (nc-SkX, blue circles and lines), including all interactions, are shown with reference to the energies of homogeneous spin spirals (dashed black line) at zero field. \textbf{d-g} Spin structure of SS, FM, SkX and isolated skyrmions on the FM background (ISk), respectively. \textbf{h-k} Spin structure of SS-2$Q$, nc-GS, nc-SkX and isolated skyrmions on the nc-GS (nc-ISk), respectively. Magnetic unit cell of nc-GS, contains three spins, is shown with solid white lines. Two spins inside the unit cell point close to the out-of-plane direction (polar angle $\theta \approx 17^{\circ}$, azimuth angle $\phi \approx 315^{\circ}$), while the third spin at the edge of the unit cell is inclined towards the in-plane direction ($\theta \approx 39^{\circ}, \phi \approx 135^{\circ}$). For better visualization, the spin direction, expect for the FM state and nc-GS, is reversed. Fourier transform of \textbf{l-n} SS, FM and SkX, respectively, excluding HOI, and \textbf{o-q} SS-2$Q$, nc-GS and nc-SkX, respectively, including HOI, in the 2DBZ. \textbf{r} Mapping of an isolated skyrmion on the noncollinear ground state (nc-ISk) onto a unit sphere, $S^2$, via stereographic projection. The color indicates the $z-$component of spins. Transition from spin up ($S_z= 1$) to spin down ($S_z= -1$) is marked by a color change from orange to dark blue via yellow.}
	\label{fig:fig1}
\end{figure*}

Here, the magnetic moment at site $i$ is denoted by $\textbf{M}_{i}$ and the corresponding unit vector is $\textbf{m}_{i}$= $\textbf{M}_{i}/\mathrm{M}_{i}$. The magnetic interaction parameters, $J_{ij}$, $\boldsymbol{D}_{ij}$ and $K$ represent the Heisenberg pairwise exchange interaction, the DMI vectors and the MAE, respectively. The external magnetic field ($B$) is applied along the $z-$direction. The HOI constants are $B_{1}$, representing the biquadratic constant, and $Y_{1}$ and  $K_{1}$ denoting the three-site and four-site four spin constant, respectively. Since these HOI arise from the fourth-order perturbation theory, we limit them to the approximation of minimal neighbor distances, as suggested in Ref.~\textsuperscript{\cite{hoffmann20}} and indicated by the summation $\langle...\rangle$.  All interaction constants were calculated via DFT for Rh/Co and Pd/Co atomic bilayers on the Re(0001) surface (see ``Methods" section for computational details and Supplementary Note 1, Supplementary Figs.~1-3 and Supplementary Tables~1-3)}.

First, we discuss the zero temperature magnetic phase diagram obtained for Rh/Co bilayers on Re(0001), excluding the HOI, i.e., the last three terms in Eq.~(1). To obtain the phase diagram, we choose three starting spin configurations: homogeneous spin spirals (SS), skyrmion lattices (SkX), and the FM state, and relax them using the atomistic spin simulations (see ``Methods" section). The energies of the relaxed states as a function of the external magnetic field are shown in Fig.~\ref{fig:fig1}b, and the relaxed spin structures are presented in Figs.~\ref{fig:fig1}d-f. 

The energies of all relaxed spin structures are displayed with reference to the energy of the homogeneous (unrelaxed) spin spiral at zero magnetic field. At zero and small magnetic fields, the relaxed spin spiral (SS) (Fig.~\ref{fig:fig1}d) becomes energetically favorable compared to the other spin structures (Fig.~\ref{fig:fig1}b). Around a magnetic field of 1.2~T, the energy of the skyrmion lattice (SkX) (Fig.~\ref{fig:fig1}f) drops below the spin spiral state and becomes the minimum energy state. With further increase of the applied magnetic field, the FM state (Fig.~\ref{fig:fig1}e) appears as the lowest energy state at around 4.1~T. This type of phase diagram is commonly observed for ultrathin TM films, such as the famous Pd/Fe atomic bilayer on the Ir(111) surface~\textsuperscript{\cite{Romming2013,Romming2015,Muckel2021}}, which possesses spin spiral minima in the energy dispersion~\textsuperscript{\cite{dupe2014,Dupe2016,malottki2017a,som18,paulhoi,paul2022}}.

The Fourier transform (FT) of the spin spiral (SS), the FM state, and the SkX are shown in Fig.~\ref{fig:fig1}l-n, respectively (see Supplementary Note 2 for details). The FT of the spin spiral state exhibits two symmetry-related peaks of equal intensity close to the $\overline{\Gamma}$ point of the 2D Brillouin zone (2DBZ), characteristic of a single-$Q$ state (Fig.~\ref{fig:fig1}l). For the FM state, the FT displays only one peak at the $\overline{\Gamma}$ point, corresponding to a vanishing wave vector, which reflects a uniform spin texture (Fig.~\ref{fig:fig1}m). The FT of the skyrmion lattice (SkX) reveals a central peak at the $\overline{\Gamma}$ point and six symmetry-related surrounding peaks of equal intensity (Fig.~\ref{fig:fig1}n). The central peak represents a uniform FM background, whereas the nearby peaks indicate a periodic modulation on the FM background in the form of skyrmions, arising from a superposition of three spin spirals related to the symmetry equivalent $\mathrm{q}$ vectors separated by 120$^\circ$.

We have also created isolated skyrmions on the FM background at and above 4.1~T according to the theoretical profile described in Ref.~\textsuperscript{\cite{bogdanov1994b}} (Fig.~\ref{fig:fig1}a) and relaxed these spin configurations using atomistic spin simulations. The relaxed isolated skyrmion (ISk) is found to be metastable on the FM background state (Fig.~\ref{fig:fig1}g). We notice that none of the spin structures, i.e., SS, SkX and ISk, change qualitatively during relaxation, leading to relatively small energy differences with respect to the zero-field homogeneous spin spiral, as evident in the phase diagram.  

Now we discuss the magnetic phase diagram of Rh/Co/Re(0001) including all interactions specified in Eq.~(1), i.e., explicitly taking into account the HOI. The obtained phase diagram (Fig.~\ref{fig:fig1}c) changes significantly from the previous one neglecting HOI (Fig.~\ref{fig:fig1}b). Here, the spin spiral is modified considerably during relaxation and transforms into a more complex spin structure (Fig.~\ref{fig:fig1}h). This relaxed spin state is a superposition of two spin spirals instead of a single (1$Q$) spin spiral state, therefore, referred to as the SS-2$Q$ state. This noticeable change in the spin orientation is associated with a large energy gain of nearly 5.7~meV/Co atom at zero magnetic field relative to the energy of the zero-field homogeneous spin spiral. As a result, the SS-2$Q$ state becomes energetically lowest at zero and low magnetic fields.

The FT of the SS-2$Q$ state (Fig.~\ref{fig:fig1}o) displays two symmetry related dominant peaks close to the $\overline{\Gamma}$ point, as obtained for spin spirals without HOI (Fig.~\ref{fig:fig1}l). Additionally, two low-intensity peaks appear at the $\pm \overline{\mathrm{K}}$ points, representing the N\'eel state
with angles of 120$^\circ$ between adjacent spins. Hence, this complex spin spiral can be interpreted as a linear superposition of two spin spirals each represented by a single-$Q$ vector, supporting its naming as the SS-2$Q$ state.

The FM starting configuration is greatly modified by the atomistic spin simulations in the presence of HOI. Surprisingly, it transforms into a noncollinear but coplanar spin configuration (Fig.~\ref{fig:fig1}i). At a critical magnetic field of about 1.7~T, this noncollinear state is energetically favorable with respect to all the other states and becomes the ground state. In the following, we refer to it as the noncollinear ground state (nc-GS). The magnetic unit cell of this noncollinear state contains three spins (Fig.~\ref{fig:fig1}i). Two of the spins located inside the 2D unit cell align close to the out-of-plane direction, whereas the spin at the boundary of the unit cell is aligned more toward the in-plane direction. The FT of the noncollinear ground state (Fig.~\ref{fig:fig1}p) displays a high intensity peak at the $\overline{\Gamma}$ point, similar to the FM state, and two more low intensity peaks at the $\pm \overline{\mathrm{K}}$ points, indicating a similar modulation as that of the SS-2$Q$ state, but on top of a uniform spin texture.

Note that the peak intensities in the FT play a crucial role in determining the spin structure. The intensities act as coefficients in the linear superposition of spin spirals, thereby influencing the final spin structure. For example, in the FT of the nc-GS, the intensity of the peak at the $\overline{\Gamma}$ point is 0.89 and at the $\pm \overline{\mathrm{K}}$ points are 0.31. A reversal of these intensities, such as 0.33 at the $\overline{\Gamma}$ point and 0.67 at the two $\overline{\mathrm{K}}$ points, yields a ferrimagnetic (FI) spin configuration as reported in Ref.~\textsuperscript{\cite{wiser2008}}.

\begin{figure*}[htbp!]
	\centering
	\includegraphics[scale=1.0]{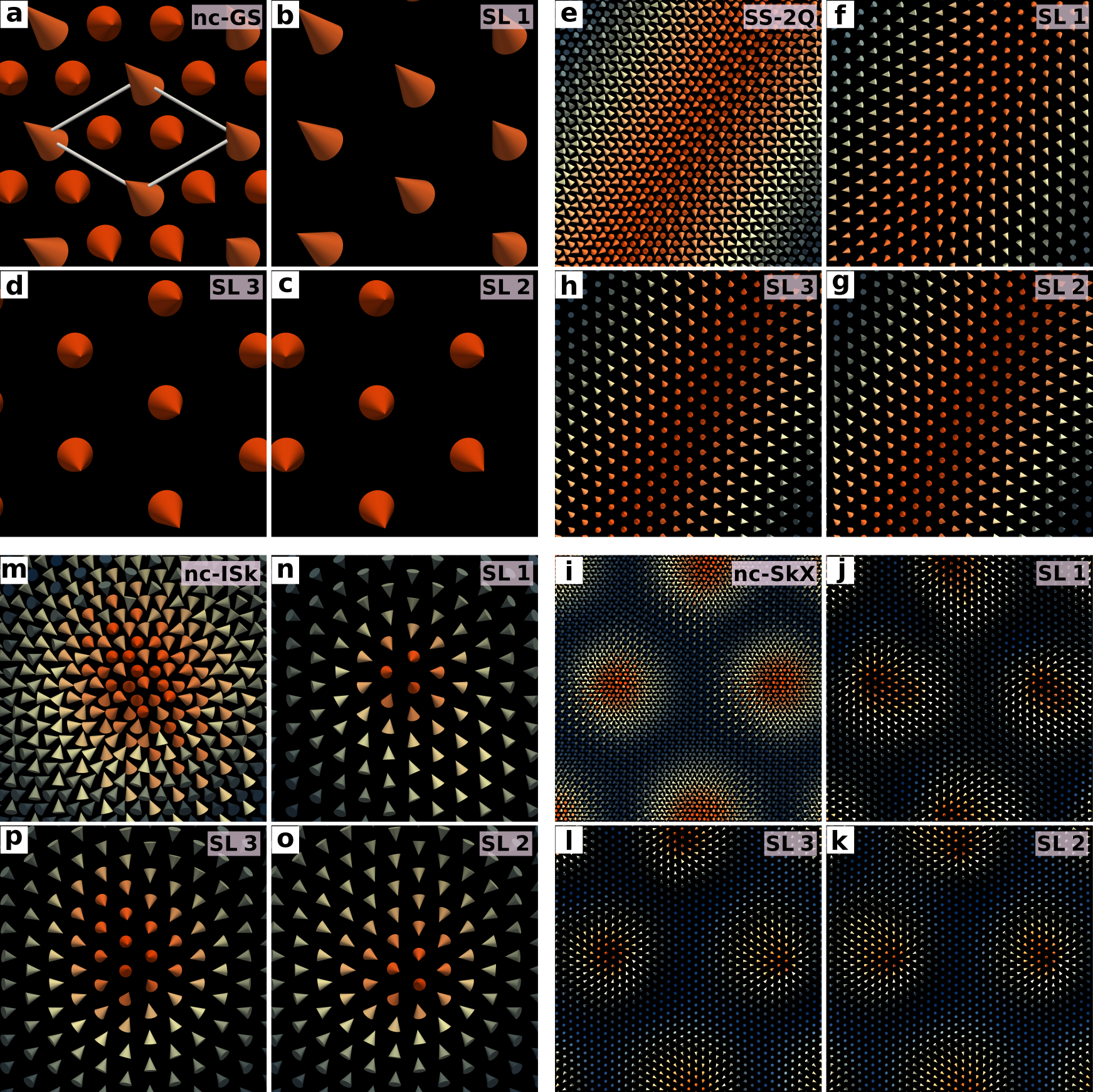}
	\caption{\justifying \textbf{Sublattice decomposition of spin structures in Rh/Co/Re(0001).} \textbf{a} The noncollinear ground state (nc-GS) in Fig.~\ref{fig:fig1}i is decomposed into \textbf{b-d} three sublattices (SL), SL~1-3. All three sublattices display FM alignment of spins, however, the magnetization axis of SL 2 (\textbf{c}) and SL 3 (\textbf{d}) points close to the out-of-plane direction ($\theta \approx 17^{\circ}$, $\phi \approx 315^{\circ}$), whereas the axis of SL 1 (\textbf{b}) is inclined more towards the in-plane direction ($\theta \approx 39^{\circ}$, $\phi \approx 135^{\circ}$). \textbf{e} Spin spiral state (SS-2$Q$) in Fig.~\ref{fig:fig1}h is decomposed into \textbf{f-h} three sublattices, SL 1-3. Each sublattice supports spin spirals characterized as a single-$Q$ state. \textbf{i} Noncollinear skyrmion lattice (nc-SkX) in Fig.~\ref{fig:fig1}j is decomposed into \textbf{j-l} three sublattices, SL 1-3. Each sublattice contains a skyrmion lattice on the FM background. However, the magnetization axes of FM background follow the same directions as of the sublattices obtained from nc-GS. The topological charge of each sublattice is same as of the composite nc-SkX state. \textbf{m} Noncollinear isolated skyrmions (nc-ISk) in Fig.~\ref{fig:fig1}k is decomposed into \textbf{n-p} three sublattices, SL 1-3. Each sublattice contains a skyrmion on the FM background. However, the magnetization axes of FM background follow the same directions as of the sublattices obtained from nc-GS. The topological charge of nc-ISk and the three sublattices is equal to 1. For better visualization, the spins of the three sublattices are rotated around an axis in the $x$-$y$ plane to align the FM spins in the out-of-plane direction. It is important to note that the rotated spin structure is nearly 6 meV per spin higher in energy than the unrotated nc-ISk spin structure.}
	\label{fig:fig2}
\end{figure*}

Using the spin model, we decompose the total energy into contributions from different interactions as a function of the angle between spins, which smoothly connects the FM and the FI state (Supplementary Fig.~4). The energy minimum occurs at around 56$^\circ$, consistent with the angle between the central and boundary spins of the nc-GS obtained via atomistic spin simulations (Fig.~\ref{fig:fig1}i). As expected, we find that the nc-GS is less favorable than the FM state in terms of pairwise exchange energy (by nearly 32 meV/Co atom) (Supplementary Table~4). However, this energy cost is compensated by the gains from both the three-site (by nearly 12 meV/Co atom) and the four-site (by nearly 26 meV/Co atom) four spin interactions, which explains the observed energy minimum. We also performed DFT calculations by varying angles and found that the total DFT energies match very well with those obtained from the atomistic spin model (see Supplementary Fig.~4).

The skyrmion lattice is also modified strikingly during relaxation when the HOI are taken into account. First, the FM background transforms into the nc-GS, and second, the spins within each skyrmion reorient themselves in such a way that the skyrmions become metastable on the nc-GS (Fig.~\ref{fig:fig1}j). We refer to this state as the noncollinear skyrmion lattice (nc-SkX). In spite of large orientation of spins during relaxation, the topological charge of the whole lattice remains unchanged, which implies that each skyrmion retains its topological identity. At a magnetic field of 1.0~T, the energy of the nc-SkX becomes lower than that of the SS-2$Q$ state (Fig.~\ref{fig:fig1}c).

In the FT of the nc-SkX state (Fig.~\ref{fig:fig1}q), we observe six symmetry-related high-intensity peaks around the $\overline{\Gamma}$ point. In addition, there are two low-intensity peaks at the $\pm \overline{\mathrm{K}}$ points. These results suggest that the HOI induce a component of spin spiral related to the $\overline{\mathrm{K}}$ point, which stabilizes the complex spin structures via creating a linear superposition with spin structures obtained excluding HOI.

To study the properties of individual skyrmions, such as radius, collapse mechanism, and energy barriers, we further relaxed isolated skyrmions on the FM background above the critical magnetic field of 1.7~T. As mentioned above, the relaxed structure contains an isolated skyrmion on the nc-GS (Fig.~\ref{fig:fig1}k), leading to a topological charge equal to one. We denote this structure as noncollinear isolated skyrmion (nc-ISk). Consequently, the spins of this structure completely warp around the unit sphere once as shown in Fig.~\ref{fig:fig1}r. Between 1.7~T and 2.2~T, these spin structures are elongated along one direction (see Supplementary Fig.~5), forming a distorted elliptical shape, and at 2.4~T, they evolve into distorted circular structures, which are shown in Fig.~\ref{fig:fig1}k.

\noindent{$\textbf{Sublattice decomposition.}$} In order to gain further insight into these complex spin structures, we divide the four complex spin structures of Fig.~\ref{fig:fig1}h-k, i.e., the SS-2$Q$, nc-GS, nc-SkX and nc-ISk, respectively, into three sublattices (SL) (Fig.~\ref{fig:fig2}). In this decomposition, the second nearest-neighbor spins in the original lattice belong to the same sublattice.

The decomposition of the nc-GS (Fig.~\ref{fig:fig2}a) shows that each SL (Fig.~\ref{fig:fig2}b-d) contains a FM state. The two spins inside the unit cell are included in SL 2 or 3, whereas the spin at the edge of the cell belongs to SL 1. Hence, the magnetization axis of the FM alignment in the SL 2 and 3 is aligned towards the out-of-plane direction, and that of SL 1 is aligned towards the in-plane direction.

In the case of the SS-2$Q$ state (Fig.~\ref{fig:fig2}e), each SL supports a spin spiral (Fig.~\ref{fig:fig2}f-h), characterized by a single-$Q$ state, but they differ in their plane of rotation. The SL 2 and 3 have a common plane of spin rotation, which differs from that of SL 1. As a result, the spin spirals in SL 2 and 3 become identical, and different from SL 1. The three SL (Fig.~\ref{fig:fig2}j-k) obtained from the decomposition of the nc-SkX (Fig.~\ref{fig:fig2}i) display a periodic arrangement of skyrmions on the same FM backgrounds (Fig.~\ref{fig:fig2}b-d) obtained from the division of the nc-GS into sublattices. Similarly, SL 2 and 3 exhibit an almost identical spin structure, and that of SL 1 is again different. Likewise, the SL 2 and 3 (Fig.~\ref{fig:fig2}o,p) obtained from the decomposition of the nc-ISk (Fig.~\ref{fig:fig2}m), host almost identical isolated skyrmions, which are distinct from those observed in SL 1 (Fig.~\ref{fig:fig2}n). Note that the individual skyrmions obtained in these SL are moderately deformed, consistent with the original lattice.

The possibility to decompose the nc-GS into three uniformly polarized sublattices lets us define a combined order parameter $\mathbf{n}=(\mathrm{R}\mathbf{m}_{\text{1}} + \mathbf{m}_{\text{2}} + \mathbf{m}_{\text{3}})/c$ with a normalization constant $c\in\mathbb{R}_{>0}$ 
and a constant rotation matrix $\mathrm{R}\in\mathrm{SO}(3)$, that rotates the canted spin in the magnetic unit cell into positive $\hat{\mathbf{z}}$-direction. This order parameter is similar to those for antiferromagnetic systems with two sublattices $\mathbf{n}'=(\mathbf{m}_{\text{1}}-\mathbf{m}_{\text{2}})/c$ and effectively assigns a vector $\mathbf{n}\in\mathbb{S}^2$ to each magnetic unit cell by taking into account the three sublattices SL~1, SL~2 and SL~3. 
Since $\mathbf{n}(\mathbf{r})=\hat{e}_z$ for $|\mathbf{r}|\to\infty$, the domain  
$\mathbb{R}^2$ is homeomorphic to $\mathbb{S}^2$ by stereographic projection 
and the existence of topological spin structures with respect to the order parameter $\mathbf{n}$ is governed by the second homotopical group $\pi_2(\mathbb{S}^2)=\mathbb{Z}$. This states, that topological textures with charge $Q\in\mathbb{Z}$ can exist within the nc-GS.

In agreement with these considerations we find a topological charge of $Q=1$ for skyrmions on each of the sublattices SL~1, SL~2 and SL~3 shown in Fig.~\ref{fig:fig2}n,p,o. With that we can assign the same charge to the collective state under the parameter $\mathbf{n}$. This construction is further supported by the fact, that we find the metastable nc-ISk 
as a local energy minimum in our atomistic spin simulations which cannot be transformed into the
nc-GS without overcoming a significant energy barrier.

We further emphasize that the possibility of obtaining distinct topological textures on a 2D lattice with a non-collinear GS is quite special. E.g.~for a triangular 2D lattice with the $120^{\circ}$
N\'eel state it is known that the order parameter is in $\mathrm{SO}(3)$. In contrast to the nc-GS in our work, the second homotopical group $\pi_2(\mathrm{SO}(3))=0$ vanishes, stating no topological protection for spin textures within the $120^{\circ}$ N\'eel state.

\begin{figure}[htbp!]
	\centering
	\includegraphics[scale=1.0]{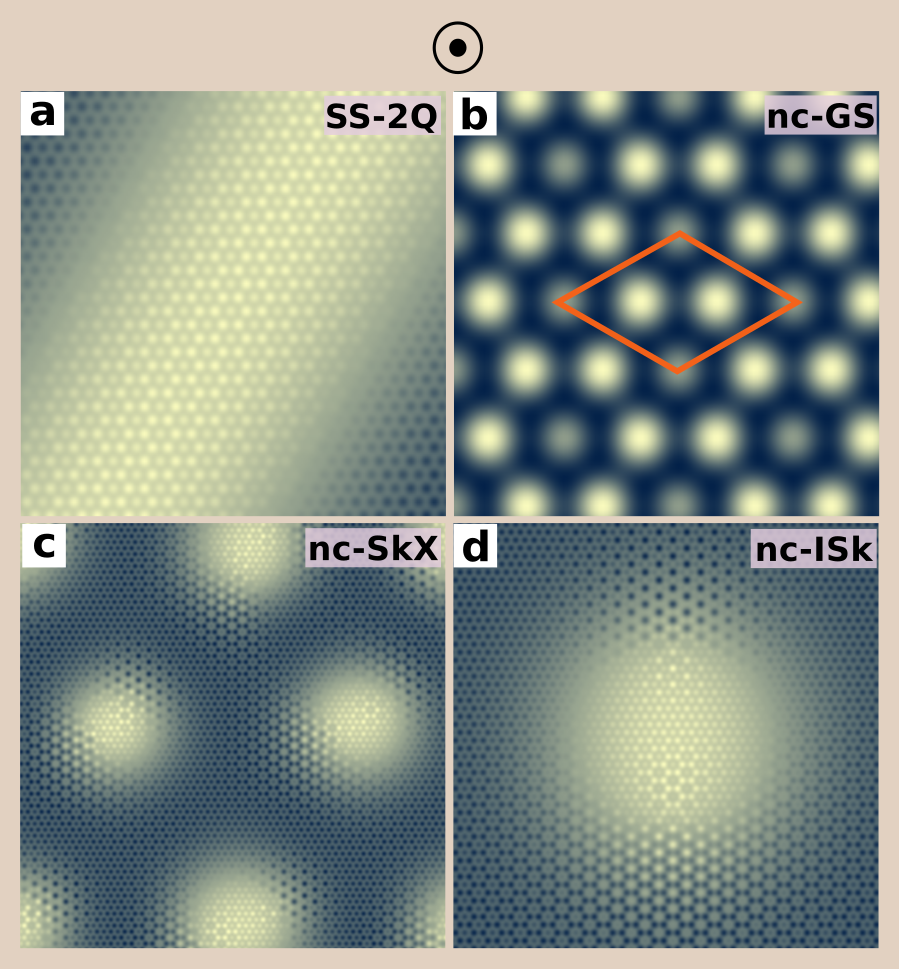}
	\caption{\justifying \textbf{Simulated SP-STM images of spin structures in Rh/Co/Re(0001).} \textbf{a-d} SP-STM image of the spin spiral (SS-2$Q$, Fig.~\ref{fig:fig1}h), the nc-GS (Fig.~\ref{fig:fig1}i), the noncollinear skyrmion lattice (nc-SkX, Fig.~\ref{fig:fig1}j), and the noncollinear isolated skyrmions (nc-ISk, Fig.~\ref{fig:fig1}k), respectively, simulated for an out-of-plane direction of the tip magnetization. The 2D magnetic unit cell of the noncollinear ground state (nc-GS) is shown with solid red lines. The tip magnetization is indicated above the panels by a circled dot for the out-of-plane direction.}
	\label{fig:fig3}
\end{figure}

\noindent{$\textbf{SP-STM images for experimental verification.}$} Since the complex spin structures we predict here exhibit noncollinear magnetic order on the atomic-scale, spin-polarized scanning tunneling microscopy (SP-STM)~\textsuperscript{\cite{wiesendanger2009}} is the ideal experimental technique to identify them. To demonstrate the feasibility, we have simulated SP-STM images~\textsuperscript{\cite{heinze2006}} (see Supplementary Note 3 for details) of the spin structures on the collinear and noncollinear backgrounds.

The SP-STM images of the spin states shown in Figs.~\ref{fig:fig1}h-k for an out-of-plane tip magnetization are shown in Fig.~\ref{fig:fig3}a-d. The image of the SS-2$Q$ state (Fig.~\ref{fig:fig3}a) displays a tilted stripe pattern, consistent with the periodicity of the spin spiral. The image of the nc-GS (Fig.~\ref{fig:fig3}b) shows a hexagonal arrangement of spins, however, the intensity of the edge spin of the magnetic unit cell is diminished relative to those that are inside the unit cell, which reflects the variation in their $z-$component of magnetization. Note that the spins sitting at the brightest spots form a honeycomb-like appearance in the simulated SP-STM image. The periodic bright spots in the image of the nc-SkX (Fig.~\ref{fig:fig3}c) show the hexagonal arrangement of skyrmions. However, the deformation in the spots indicates noncircular skyrmion profiles and the atomic-scale contrast of the nc-GS is visible. In the SP-STM image of noncollinear isolated skyrmions (Fig.~\ref{fig:fig3}d), the internal noncollinearity of the spin structure is also visible.

\begin{figure*}[htbp!]
	\centering
	\includegraphics[scale=1.0]{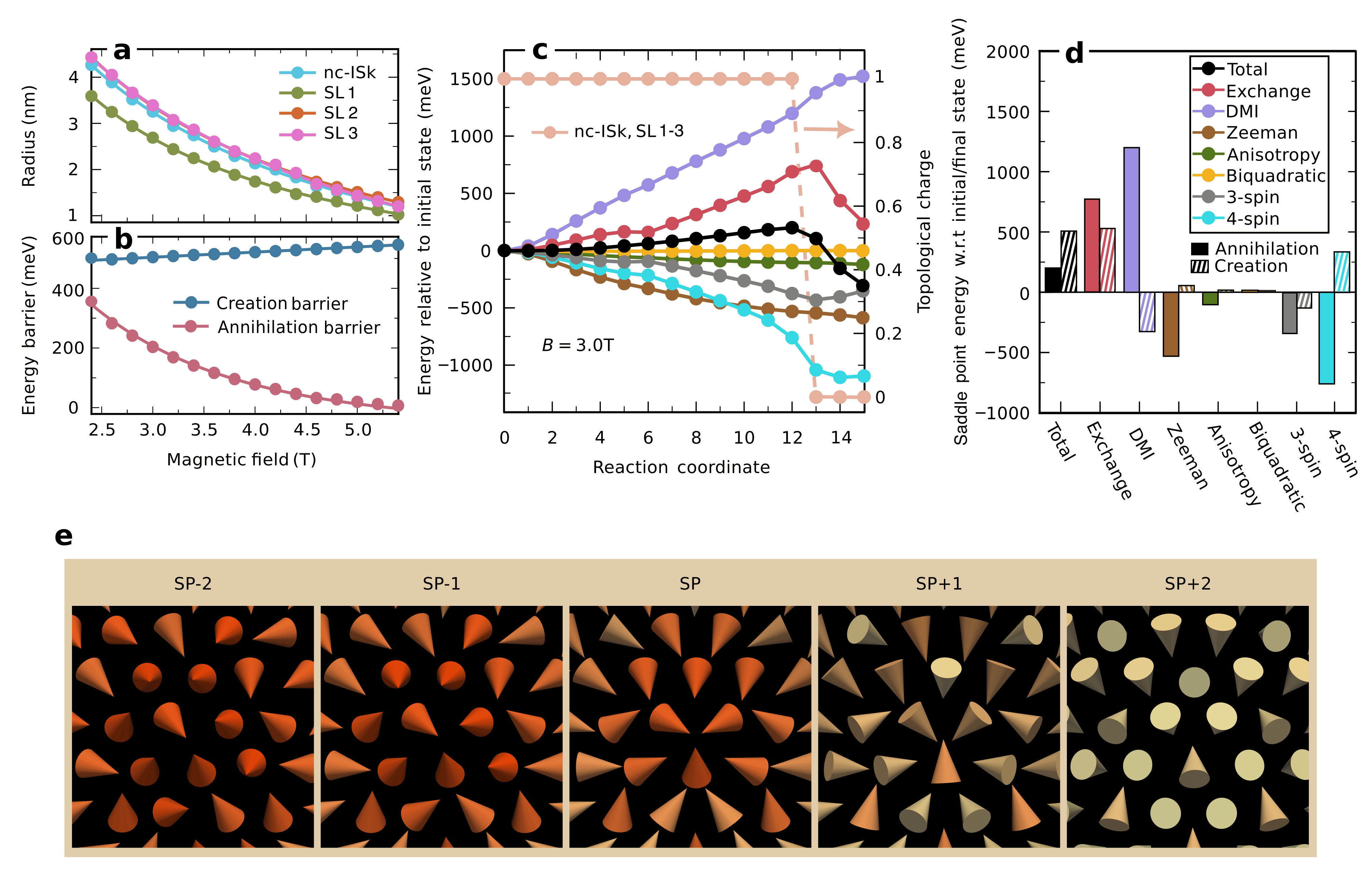}
	\caption{\justifying \textbf{Radius, minimum energy path, and energy barriers for skyrmions in Rh/Co/Re(0001).} \textbf{a} Radius of noncollinear isolated skyrmions (nc-ISk) in the original lattice (Fig.~\ref{fig:fig2}m) and three sublattices (Fig.~\ref{fig:fig2}n-p) as a function of the applied magnetic field. The filled circles represent data from the atomistic spin simulations, the solid lines are fits to the function $y(x)= a \left( x+b \right)^{-1} + c$, where $y$ is the skyrmions radius, $x$ is the magnetic field, and $a$, $b$ and $c$ are fitting constants. \textbf{b} Creation and annihilation energy barriers of the noncollinear isolated skyrmions (nc-ISk) as a function of applied magnetic field. The atomistic spin simulation data (filled circles) of the annihilation and creation barriers are fitted with $y(x)= a \left( x+b \right)^{-1} + c$ and $y(x)= mx + c$, respectively, where $y$ is the energy barriers, $x$ is the magnetic field, and $a$, $b$, $c$, and $m$ are fitting constants. \textbf{c} Total and individual energy contributions along the minimum energy path (MEP) corresponding to the radial collapse of the noncollinear isolated skyrmions (nc-ISk) at 3.0 T including all interactions. The energy of the saddle point relative to the initial state (nc-ISk) represents the annihilation barriers, whereas its energy relative to the final state (nc-GS) represents the creation barriers. Topological charge of noncollinear isolated skyrmions (nc-ISk) and isolated skyrmions on the FM background in three sublattices along MEP is also shown. \textbf{d} Energy decomposition of the saddle point with respect to the initial (nc-ISk) and final states (nc-GS). \textbf{e} Spin structures of noncollinear isolated skyrmions (nc-ISk) at the saddle point (SP), before the saddle point (SP-2 and SP-1) and after the saddle point (SP+1 and SP+2) corresponding to the minimum energy path in \textbf{c}. The three-site ($Y_1$) four spin interaction and four-site four spin interactions ($K_1$) are abbreviated as 3-spin and 4-spin, respectively. The values of the fitting constants are listed in Supplementary Table~5.}
	\label{fig:fig4}
\end{figure*}

We further calculated the SP-STM images of the spin states in Figs.~\ref{fig:fig1}h-k for tip magnetization along the two in-plane directions (Supplementary Fig.~6a-h). While the SS-2$Q$ shows similar patterns for both in-plane and out-of-plane tip magnetizations, the nc-GS displays distinct patterns compared to those obtained with the out-of-plane tip magnetization. The dark and light contrast of the noncollinear skyrmion lattice and noncollinear isolated skyrmions with in-plane tip magnetization is absent for an out-of-plane tip magnetization. In all cases, the noncollinear background induced by the higher-order exchange interactions can be clearly identified, suggesting the feasibility of experimental verification. 

Due to the noncollinear background, the SP-STM images (Fig.~\ref{fig:fig3}a-d and Supplementary Fig.~6a-h) look qualitatively different from those obtained for spin structures neglecting HOI (Supplementary Fig.~7a-j). The latter SP-STM images agree with those observed experimentally for magnetic skyrmions in Pd/Fe/Ir(111)~\textsuperscript{\cite{Romming2013,Romming2015}}. Thus, we conclude that the proposed unconventional isolated skyrmions and skyrmion lattices on the noncollinear background can be clearly identified via SP-STM. 

\noindent{$\textbf{Size and stability of isolated skyrmions.}$} Next we focus on the properties of noncollinear isolated skyrmions (nc-ISk), such as, radius, collapse mechanism and energy barriers. Due to their complex spin structures, the theoretical skyrmion profile derived in Ref.~\textsuperscript{\cite{bogdanov1994b}} cannot be used to compute their radius. Therefore, we evaluated it by taking a closed contour around the isolated skyrmions (see Supplementary Note 4 for details) in the original lattice (Fig.~\ref{fig:fig2}m) as well as on the three sublattices (Fig.~\ref{fig:fig2}n-p). 

Above the critical magnetic field of 1.7~T (Fig.~\ref{fig:fig1}c), the isolated skyrmions are metastable on the noncollinear background. In the field range of 1.7 to 2.2~T, however, these structures are elongated (see Supplementary Fig.~5). At a field of 2.4~T, they transform into relatively circular shaped spin structures, and hence, we choose to study their properties from this field value onward.

The radius of the isolated skyrmions in the original lattice is 4.3 nm at an applied magnetic field of 2.4~T, and it decreases to a value of 1.3~nm at 5.4~T (Fig.~\ref{fig:fig4}a). These skyrmions are about twice larger in diameter than skyrmions observed experimentally in Rh/Co~\textsuperscript{\cite{meyer19}} and Pd/Fe atomic bilayers on Ir(111)~\textsuperscript{\cite{Romming2015,Muckel2021}}. The radius of isolated skyrmions in the sublattices also varies in a similar manner with the magnetic field (Fig.~\ref{fig:fig4}a). Skyrmions in sublattice 2 and 3 (Fig.~\ref{fig:fig2}o,p) exhibit similar radii, due to their equivalent spin textures, and their radii closely match to the skyrmions in the original lattice, while those in sublattice 1 (Fig.~\ref{fig:fig2}n) display a slightly smaller radius. 

To assess their stability, we calculated the collapse of isolated skyrmions (nc-ISk) into the noncollinear ground state (nc-GS) using the geodesic nudged elastic band (GNEB) method~\textsuperscript{\cite{bessarab2015}} (see ``Methods") and evaluated the minimum energy path (MEP), which allows us to identify the transition mechanism. Despite the noncollinear background, we found that these unconventional skyrmions follow the same radial collapse mechanism for annihilation, as seen for the conventional skyrmions~\textsuperscript{\cite{Muckel2021}} (Supplementary Fig.~8). The isolated skyrmions in the three sublattices (Fig.~\ref{fig:fig2}n-p) also annihilate through the radial collapse mechanism (Supplementary Figs.~9-11).

The energy of the saddle point, i.e., the point of highest energy along the MEP, relative to the initial state (nc-ISk) and final state (nc-GS) defines the annihilation and creation barriers, respectively. The annihilation energy barriers, which prevent these skyrmions from collapsing, vary in a similar way with the magnetic field as the radius (Fig.~\ref{fig:fig4}b). The energy barrier is 335 meV at a magnetic field of 2.4~T and it decreases to a value of 6.8 meV at 5.4~T. This behavior can be understood from the fact that all the energy terms that contribute to the energy barrier scale with skyrmion size. The calculated energy barriers for skyrmions in Pd/Fe/Ir(111)~\textsuperscript{\cite{paulhoi}} and Rh/Co/Ir(111)~\textsuperscript{\cite{meyer19}}, which were experimentally observed using SP-STM, are around 240 and 280 meV, respectively. These values suggest that the unconventional skyrmions predicted here can be realized in experiments. In contrast, the creation barriers show a very small linear increment with the magnetic field, which was observed previously in Ref.~\textsuperscript{\cite{paul2022}}. This linear behavior arises from a competition between the annihilation barriers and the energy difference between the initial and final states.

To study the collapse mechanism, we present the energy decomposition of the MEP at an applied magnetic field of 3~T (Fig.~\ref{fig:fig4}c). The DMI clearly favors the nc-ISk state with respect to the nc-GS. This energy term scales with the number of spins in the skyrmion. Therefore, as the size of the nc-ISk shrinks along the MEP, the DMI energy increases linearly. The Zeeman energy, in contrast, favors the nc-GS over the nc-ISk state and decreases linearly in energy along the MEP. The exchange interaction energy rises up to the saddle point as the relative angles between spins increase, and then decreases beyond this critical point as the spins realign with the nc-GS (Fig.~\ref{fig:fig4}e). This term favors the nc-ISk state over the nc-GS, and destabilizes the latter. The three-site four spin term exhibits the opposite behavior to the exchange term. The four-site four spin interaction, which along with the three-site term stabilizes the nc-GS, decreases linearly and undergoes a sharp drop due to the large reorientation of spins around the saddle point, and remains almost constant thereafter due to the minor change in spin alignment (Fig.~\ref{fig:fig4}e). The calculation of the topological charge of the isolated skyrmions in the main lattice and in each sublattice along MEP indicates that they remain topologically protected up to the saddle point, and this protection breaks beyond this critical configuration (Fig.~\ref{fig:fig4}c).

The decomposition of the energy barrier into the contributions of individual interactions reveals that the DMI and exchange interactions contribute with positive values to the annihilation barrier, while the three-site four spin interaction, the four-site four spin interaction, the MAE, and the Zeeman term contribute with negative values, since they favor the nc-GS (Fig.~\ref{fig:fig4}d). The contribution of these terms changes sign for the creation barrier, except for the exchange interaction term, which remains positive, consistent with the behavior of the individual terms described above.

\begin{figure}[htbp!]
	\includegraphics[scale=1.0]{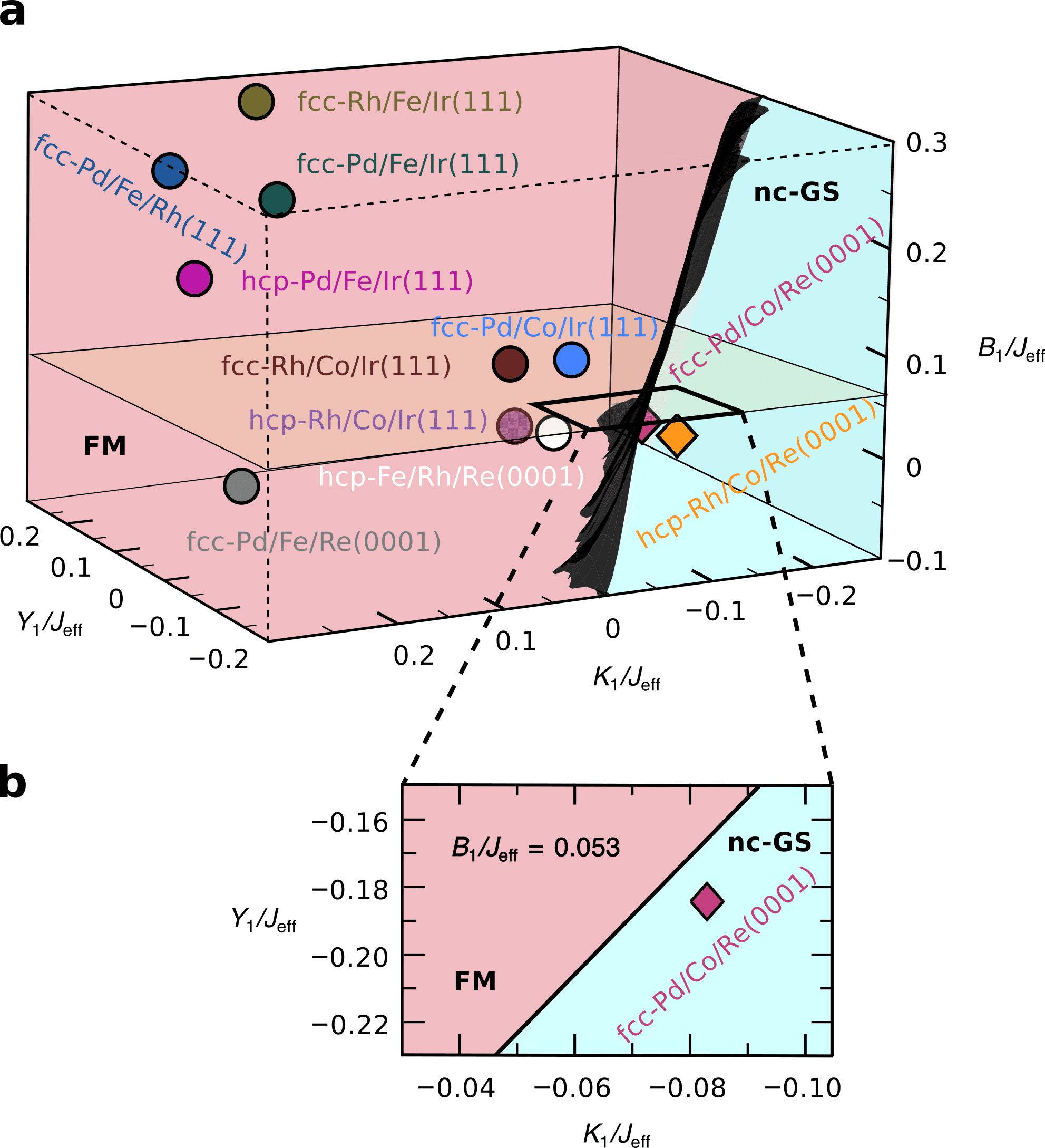}
	\caption{\justifying \textbf{Ground states from a competition of HOI 
    and pair-wise exchange.} \textbf{a} Phase diagram displaying the variation of three higher-order exchange interactions, i.e., three-site four spin ($Y_1$), four-site four spin ($K_1$) and biquadratic ($B_1$) interactions, as a function of effective exchange interaction ($J_{\mathrm{eff}}$). The light red indicates the FM phase, while the cyan denotes the noncollinear phase as shown in Fig.~\ref{fig:fig1}i, on which the isolated skyrmions are formed. The black surface marks the boundary of these two phases. The filled circles present previously studied TM ultrathin films~\textsuperscript{\cite{malottki2017a,meyer19,paul2020b,paulhoi,romming18,gutzeit2021,meyer2020}} and the filled diamonds indicate the two films studied here. The magnetic interactions of the additional TM ultrathin films used here are summarized in Supplementary Table~6. \textbf{b} Zoomed view around fcc-Pd/Co/Re(0001) at $B_1/J_{\mathrm{eff}}=$ 0.053. The effect of DMI ($D_{\mathrm{eff}}$) and MAE ($K$) on the phase diagram is negligible (Supplementary Fig.~16).}
	\label{fig:fig5}
\end{figure}

\noindent{$\textbf{Discussion.}$} We have demonstrated that the FM state of Rh/Co bilayers on the Re(0001) surface is destabilized by higher-order interactions. As a result, a noncollinear ground state, which has not been reported previously, becomes energetically favorable for Rh/Co/Re(0001) at an external magnetic field, and unconventional isolated skyrmions are metastable on this noncollinear ground state due to the effect of DMI and pairwise exchange interaction. The decomposition of the main lattice into three spin sublattices reveals skyrmions on a FM state with different magnetization axes on each sublattice. The energy barriers protecting these unconventional skyrmions from collapsing into the noncollinear state are comparable to those conventional skyrmions previously reported in Pd/Fe/Ir(111)~\textsuperscript{\cite{malottki2017a,paulhoi}}, which suggests that they can be observed in the experiment. 

For Pd/Co bilayers on the Re(0001) surface, we have found the emergence of the same noncollinear magnetic ground state due to higher-order exchange interactions (see Supplementary Note~5 and Supplementary Figs.~12-15). In this system, the noncollinear state becomes the ground state even at zero magnetic field and isolated skyrmions are metastable on this noncollinear state at zero field. This demonstrates that the possible destabilization of the FM state can occur more generally, and, quite surprisingly, even in films consisting of the prototypical strong ferromagnetic material Co.

Upon closer inspection, we notice that, in the two studied film systems, two types of four-spin interaction terms dominate over the biquadratic term (Supplementary Tables~1-3). In contrast to previously studied Co-based film systems, in which isolated skyrmions have been reported on the FM background, both of these four-spin interactions exhibit a negative sign~\textsuperscript{\cite{gutzeit2021}}. These observations raise a crucial question: which HOI term is primarily responsible for stabilizing the noncollinear ground state in these two films, and more specifically, what sign and magnitude of these terms relative to the pairwise exchange term -- which favors the FM state -- are required to stabilize the noncollinear ground state?

To address these issues, we calculated the phase diagram (Fig.~\ref{fig:fig5}) by varying the three HOI, i.e., the biquadratic ($B_1$), the three-site ($Y_1$) and the four-site ($K_1$) four spin interactions, independently as a function of the effective exchange interaction ($J_{\mathrm{eff}}$). The effective exchange constant $J_{\mathrm{eff}}$ is equal to the exchange constant $J_1$ when only the nearest-neighbor exchange is present in the film system. In a real system, the effective exchange constant is determined from a parabolic fit to the energy dispersion of spin spirals in the vicinity of the FM state ($\overline{\Gamma}$ point) and thereby, includes the effect of exchange frustration, i.e., terms beyond the nearest-neighbor~\textsuperscript{\cite{Dupe2016,meyer19}}. 

The obtained phase diagram (Fig.~\ref{fig:fig5}) displays two distinct regions: one corresponding to the FM ground state and another region corresponding to the noncollinear ground state (nc-GS). The latter state extends over a considerable part of the phase diagram within a moderate range of the three HOI constants relative to the effective (pairwise) exchange interaction, while the FM state extends over a larger portion of the phase diagram and arises outside this range of the HOI constants. It is apparent from the phase diagram that the negative values of the three HOI constants favor the noncollinear ground state, since the phase diagram is simulated with positive effective exchange constants only.

The two Co-based ultrathin film systems studied in this work belong to the noncollinear part of the phase diagram (Fig.~\ref{fig:fig5}a,b), whereas all other previously studied films~\textsuperscript{\cite{malottki2017a,meyer19,paul2020b,paulhoi,romming18,gutzeit2021,meyer2020}} remain in the FM region (see Supplementary Table~6 for the interaction constants). Interestingly, the Co-based film systems exhibit negative values of the three-site and four-site four spin interaction constants, while positive interaction constants are typical for Fe films (see Supplementary Table~6). The values indicate that, for the two films studied here, the contribution of the biquadratic term is not required to stabilize the noncollinear state, and the other two four spin interactions play the key role for the stability. However, this finding is not true for the entire part of the phase diagram. When the strength of the biquadratic term becomes relatively stronger, it also plays a noteworthy role in stabilizing the noncollinear ground state.

We further find that the angle between the spins within the magnetic unit cell of the noncollinear state changes for the two films studied here. It is 56$^\circ$ for Rh/Co/Re(0001) and 34$^\circ$ for Pd/Co/Re(0001). In general, the angle depends on the strength of the HOI constants relative to the effective exchange constant (see Supplementary Fig.~17). Note that the phase boundary separating the nc-GS and the FM state remains unchanged upon including the DMI and MAE terms (Supplementary Fig.~16). These findings show that the stability of the complex noncollinear ground state, on which the unconventional isolated skyrmions can emerge spontaneously, depends on the competition between the pairwise exchange and higher-order exchange interactions.

Our analysis reveals that the strength and sign of the HOI compared to the pairwise exchange interaction are crucial for the destabilization of the FM state, leading to the stabilization of the complex noncollinear ground state in the two films studied here. However, the phase diagram clearly displays that this destabilization mechanism is not limited to these two films alone. It can be extended to other materials which have the desired higher-order to exchange interaction ratio, such as TM based ultrathin films or 2D van der Waals (vdW) magnets, in which HOI can play a significant role. With typical strengths of the other interactions specified in the extended spin model, such as the DMI, MAE and Zeeman, the unconventional skyrmions can emerge on the complex noncollinear state. This insight indicates the possibility of a broader class of promising materials in which this new type of complex spin structure could be realized and observed experimentally.\\

\noindent{\large{\textbf{\MakeUppercase{Methods}}}}\\
\noindent\textbf{Density functional theory.} The density functional theory (DFT) calculations of the
energy dispersion of spin spirals without SOC, the magnetic moments, and the MAE constant for hcp-Rh and fcc-Pd monolayers on Co/Re(0001) have already been presented in Ref.~\textsuperscript{\cite{paul2024}}. Here, we computed the DMI and HOI constants for the two film systems. The DMI energy is evaluated within first-order perturbation theory~\textsuperscript{\cite{heide2009}} based on the self-consistent spin spiral states obtained previously~\textsuperscript{\cite{paul2024}} using the \textsc{Fleur} code~\textsuperscript{\cite{fleur}}. Details of calculating the three HOI constants are described in Ref.~\textsuperscript{\cite{paulhoi}}. Here, we evaluated the HOI terms using the Vienna ab initio simulation package (VASP) code~\textsuperscript{\cite{vasp,Kresse1996}}, which is based on the projected augmented wave (PAW) formalism~\textsuperscript{\cite{paw,Kresse1999}}. We have seen that these two DFT codes produce results in good agreement with each other~\textsuperscript{\cite{paulhoi,paul2020b,paul2022,li2020}}. We used the same exchange-correlation functional and DFT computational parameters as in Ref.~\textsuperscript{\cite{paul2024}} for DMI calculations and as in Ref.~\textsuperscript{\cite{paulhoi}} for the evaluation of the three HOI constants. \\

\noindent\textbf{Atomistic spin simulations.} To relax the spin structures and accurately find the energy minima, we used a robust approach by combining two methods. We started the relaxation with the Landau-Lifshitz dynamics, however, as the system approaches the energy minima, we shifted to the velocity projection optimization (VPO) method~\textsuperscript{\cite{bessarab2015}}.

The relaxation of a system of classical spins is governed by the Landau-Lifshitz equation
\begin{gather} \label{eq:llg}
\hbar \frac{d\textbf{m}_{i}}{dt}=\frac{\partial \mathcal{H}}{\partial{\textbf{m}}_{i}}\times\textbf{m}_{i}-\alpha\left(\frac{\partial \mathcal{H}}{\partial{\textbf{m}}_{i}}\times\textbf{m}_{i}\right)\times\textbf{m}_{i}
\end{gather}
where $\hbar$ is the reduced Planck constant, $\textbf{m}_{i}$ is the unit vector of the magnetic moment at site $i$, $\alpha$ is the damping parameter and the Hamiltonian  $\mathcal{H}$ defined in Eq.~(1). The time integration of the Landau-Lifshitz equation is performed via the semi-implicit scheme~\textsuperscript{\cite{mentink2000}} proposed by Mentink $\textit{et al.}$ A time step of 0.1 fs along with a value of $\alpha$ between 0.25 and 0.50 is used to relax the spin structures for 1 million steps.

In the VPO method, which basically resembles a Velocity-Verlet algorithm, only the gradient is taken into account for iterate updates of the spins towards the closest local minimum configuration. Thus, in contrast to the Landau-Lifshitz method, the spin space trajectory is not physical, but it allows for numerical speed-up techniques, such as restricting the velocities to components pointing along the direction of the gradient only. This projection scheme enhances the efficiency of this method even in a flat landscape, where the Landau-Lifshitz method becomes stagnant~\textsuperscript{\cite{ivanov2021fast}}. 

All results presented in the main text and Supplementary information (Supplementary Figs.~4-17) are simulated using a hexagonal lattice of 150$\times$150 spins. However, we obtained the same spin structures and results using a lattice of 180x180 spins (see Supplementary Fig.~18).\\

\noindent\textbf{Geodesic nudged elastic band method.} We calculated the minimum energy path (MEP) between the noncollinear isolated skyrmion (nc-ISk) and noncollinear ground state (nc-GS) using the geodesic nudged elastic band (GNEB) method~\textsuperscript{\cite{bessarab2015}}. 

In this approach, we initialize a discrete path connecting the initial (nc-ISk) and final (nc-GS) states by intermediate spin configurations, referred to as images, separated by spring forces. The aim of this method is to optimize the energy along the path through iterative relaxation of the images resulting in the MEP. The relaxation is performed through a force projection scheme, where the perpendicular component of the effective field, calculated at each image, is kept, and the tangential component is replaced by the spring force to maintain a uniform distribution of images along the path. The maximum energy along the MEP corresponds to the saddle point (SP), which defines the annihilation and creation barriers with respect to the initial and final states, respectively. The climbing image (CI) technique is used on top of the GNEB method to accurately measure the energy of the saddle point.\\
   
\noindent{\textbf{\MakeUppercase{Data availability}}\\ 
\noindent{The authors declare that the data supporting the findings of this study are available within the article and its Supplementary Information file.}\\
	
\noindent{\textbf{\MakeUppercase{Code availability}}\\
\noindent{The atomistic spin simulation code, Spinaker, used in this study is available from the authors upon reasonable request.}\\

\noindent{\textbf{\MakeUppercase{References}}\\
\bibliographystyle{naturemag}

\vspace{0.05in}
\noindent{\textbf{\MakeUppercase{Acknowledgments}}\\ 
S.P. acknowledges funding from the Anusandhan National Research Foundation (ANRF/ECRG/2024/001865/PMS). S.H. thanks the Deutsche Forschungsgemeinschaft for funding via SPP2137 “Skyrmionics” (project no.~462602351) and  project no.~418425860. The authors gratefully acknowledge the computing time made available to them on the high-performance computers ``Lise" at the NHR center NHR@ZIB and ``Emmy" at the NHR center NHR@G\"{o}ttingen.  These centers are jointly supported by the Federal Ministry of Education and Research and the state governments participating in the NHR~\textsuperscript{\cite{nhr}}. S.P. thanks IISER Thiruvananthapuram for funding and computing time on the Padmanabha cluster. H.S. acknowledges financial support from the Icelandic Research Fund (grant No. 239435). We thank Tim Drevelow for valuable discussions.\\
	
\noindent{\textbf{\MakeUppercase{Author contributions}}\\
S.P. conceived and supervised the project, and performed the DFT calculations. M.B. and M.G. performed the atomistic spin simulations using the Spinaker code with the help of M.A.G., B.B., and H.S. M.G. and M.B. prepared the figures. S.P. and S.H. wrote the paper. All authors discussed and analyzed the results and contributed to the preparation of the paper. \\
	
\noindent{\textbf{\MakeUppercase{Funding}}\\
\noindent{Open access funding enabled and organized by Projekt DEAL.}\\	
	
\noindent{\textbf{\MakeUppercase{Competing interests}}\\ The authors declare no competing financial or non-financial interests. \\
		
\noindent{\textbf{\MakeUppercase{Additional information}} Supplementary Information accompanies this paper.
				
\end{document}